\documentclass[conference]{IEEEtran}
\usepackage{amsfonts}
\usepackage{amsmath}
\usepackage{graphicx, graphics,epsfig,subfigure}
\usepackage{color, cite}
\usepackage{amssymb, amsmath}
\usepackage{algorithm}
\usepackage{algorithmic}
\usepackage{url}
\usepackage{amsfonts, dsfont}
\usepackage{latexsym}
\usepackage{tabularx}
\usepackage{comment}
\usepackage{colortbl}

\newcommand{\beq}   {\begin{equation}}
\newcommand{\eeq}   {\end{equation}}
\newcommand{\bep}   {\begin{equation*}}
\newcommand{\eep}   {\end{equation*}}
\newcommand{\bea}   {\begin{eqnarray}}
\newcommand{\eea}   {\end{eqnarray}}
\newcommand{\bda}   {\begin{eqnarray*}}
\newcommand{\eda}   {\end{eqnarray*}}
\newcommand{\bdalign}   {\begin{align*}}
\newcommand{\edalign}   {\end{align*}}

\newtheorem{theorem}{Theorem}
\newtheorem{conjecture}[theorem]{Conjecture}

\newtheorem{definition}[theorem]{Definition}

\begin{document}

\title{On Stability and Sojourn Time of Peer-to-Peer Queuing Systems}
\author{
Taoyu Li$^*$ ~ Minghua Chen$^\dagger$ ~ Tony Lee$^\dagger$ ~ Xing Li$^*$ \\
$^*$Tsinghua University, Beijing, China. {\{ldy03@mails.tsinghua.edu.cn,xing@cernet.edu.cn\}} \\
$^\dagger$The Chinese University of Hong Kong, Shatin, Hong Kong. {\{minghua,ttlee\}@ie.cuhk.edu.hk}
}

\maketitle

\begin{abstract}
Recent development of peer-to-peer (P2P) services
systems introduces a new type of queue
systems that receive little attention before, where both job and server arrive and
depart randomly. Current study on these models focuses on the stability
condition, under exponential workload assumption.
This paper extends existing result in two aspects.
In the first part of the paper we relax the exponential workload assumption,
and study the stability of systems with general workload distribution.
The second part of the paper focuses on the job sojourn time.
An upper bound and a lower bound for job sojourn time are investigated.
We evaluate tightness of the bounds by numerical analysis.
%
%

\end{abstract}



\section{Introduction}\label{sec:Intro}

Classical queueing theory have been a useful tool on modeling systems
where jobs arrive randomly at static service stations of given service capacities.
It provides analysis of the system's properties, such as stability and sojourn time.
Queueing theory has wide application in many scenarios of operations research.
In particular, its application in studying computer networks and operating systems led to a
generalization of queueing theory to model a network of queues and
many different service policies \cite{kleinrock1975qs.v1, baskett1975oca}.

Recently, the modeling of peer-to-peer (P2P) systems is pointing to
a new kind of queueing system. In this new model,
jobs still arrive randomly, but the service stations also arrive
randomly \cite{2009iptps}. This new model, if well studied, would be
helpful on the quality of service study of different kinds of
peer-to-peer systems, such as online storage or video-on-demand
systems\cite{zhang2009isip}.

However, current study on the performance of this new model has been
limited on the system stability condition only, and under
exponential workload assumptions. In this paper we extend
this result on two aspects. We first relax the exponential
workload assumption, and prove that a P2P
queuing system with general workload distribution would still
have the same stability condition as the system with exponential workload. Then we
turn ourselves go beyond stability and study the sojourn time of the p2p queuing systems under the exponential workload assumption.
Such sojourn time is challenging to derive.
We provide an upper bound and a lower bound for it, and show the bound is tight by numerical analysis.

The rest of the paper is organized as followed:
We first brief introduce the related work on peer-to-peer system modeling and queueing model with dynamic service rate in section~\ref{sec:related},
and list the basic notations and assumptions in section~\ref{sec:model}.
Then in section~\ref{sec:mgmmstable} we study the system stability without exponential assumption.
Upper and lower bounds of average queue length is derived in section~\ref{sec:mmmmenc}.
Section~\ref{sec:simulation} gives some numerical results.
The work is concluded in section~\ref{sec:concl}.

%
%
%

\section{Related Work}
\label{sec:related}

The modeling of P2P systems have received a lot of attention from researchers in the past years.
One of the first dynamic P2P models was introduced by Qiu and
Srikant \cite{qiu2004map} to model BitTorrent, a P2P file sharing
system. The model is simple, but very inspiring. Although they did
not mention queueing theory, they implicitly modeled randomly
arriving service stations (which are peers themselves) providing an
\emph{effective} service (file sharing) rate.  Subsequently, Fan,
Chiu and Lui \cite{fan2006dtb} modeled and studied the tradeoff
between different service rate allocations in a dynamic P2P system
similar to the Qiu-Srikant model.

Clevenot, Nain and Ross \cite{ross2005pe} generalized the
Qiu-Srikant model fluid model to describe more realistic cases. The
authors in \cite{kumar2007infocom} and \cite{wu2009infocom} also
used randomly arriving peers with certain service rate to model P2P
streaming systems.

On the other hand, the queuing model with dynamic service have been preliminarily studied in other areas,
such as in modeling transportation system \cite{melike2004qs}.
There are also research on performance of a queuing system with varying service rate.
However, these works lies in two extreme cases.
Some of them focus on special case that there is only two state of service rate:
``normal'' and ``disrupted''\cite{melike2004qs,keilson1993aap}.
While rests study models which are too general that they can only give numerical results
\cite{sai2005qs,neuts1979ms,mitrany1968or}.
Both of them are not suitable to study P2P systems, where service rate may varies proportionally with the number of available servers in the system.
Our previous work\cite{2009iptps} models this kind of systems, and gives a preliminary study.

\section{Queuing Model for Peer-to-Peer Systems} \label{sec:model}

This paper is based on the framework of P2P queuing model proposed
in our previous work\cite{2009iptps}. The notations used in the
framework are listed in Table~\ref{tab:notations}.

\begin{table}[!htp]
\caption{Key Notation}\label{tab:notations} \centering
\begin{tabular}
[l]{l|l} \hline
\textbf{Notation} &  \textbf{Definition} \\
\hline
A & job arrival process.\\
B & job service time distribution.\\
s & number of servers (used in traditional queuing model).\\
C & server arrival process.\\
E & server life time distribution.\\
\hline
$1/\lambda_c$ & average interarrival time between two job arrivals. \\
$1/\mu_c$ & average job service time when served by a single server. \\
$\rho_c=\lambda_c /\mu_c$ & job load demand of the system.\\
$1/\lambda_s$ & average interarrival time between two server arrivals. \\
$1/\mu_s$ & average server life time. \\
$\rho_s=\lambda_s /\mu_s$ & service capacity of the system.\\
$n_c(t)$ & the number of jobs in the system at time $t$. \\
$n_s(t)$ & the number of servers in the system at time $t$. \\
$X(t)$ & total job service time in the system at time $t$ when served\\
&  by a single server.\\
 \hline
\end{tabular}
\end{table}

In classical queuing theory, Kendall's
notation~\cite{kleinrock1975qs.v1}, i.e.,
$\mbox{A}/\mbox{B}/\mbox{s}$, is widely used to
represent queuing model for static service systems where servers are
static.

To represent new queuing models for P2P service systems in which
both job and server dynamically arrive and depart, we use the notation
\begin{equation}\label{eq:model}
\mbox{A}/\mbox{B}/(\mbox{C}/\mbox{E}).
\end{equation}
This notation extends Kendall's notation, by using two additional
terms (C and E) to represent the server dynamics in P2P queuing
models.

To represent systems with different job and server dynamics, some
notations we use for arrival processes (A or C) and distributions (B
or E) are 1) $M$ for a memoryless process (e.g. Poisson process), or
an exponential distribution. 2) $D$ for a
deterministic process, or a deterministic distribution. 3) $G$ for a
process with general independent arrivals, or arbitrary
distribution.

In this paper we mainly discuss about $M/M/(M/M)$ and $M/G/(M/M)$ systems, with two basic assumptions:
\begin{enumerate}
\item Servers are homogeneous and each has a unit service capacity.
\item Job arrival are independent to the server process.
\end{enumerate}

\section{System Stability}
\label{sec:mgmmstable}

Our previous work\cite{2009iptps} has discussed the stability of an $M/M/(M/M)$ queuing system,
in the semantic in the positive recurrence of the job process,
resulting in Theorem.\ref{thm:mmmm.Stability}.

\begin{definition}[\textbf{Stability}]\label{def:stability}
A P2P service system is stable if its corresponding job-server
process $\left\{ n_c(t),n_{s}(t)\right\} _{t}$ or $\left\{ X(t),n_{s}(t)\right\} _{t}$ is positive
recurrent, so that a stationary distribution exists.
\end{definition}

\begin{theorem}[\textbf{Stability for $M/M/(M/M)$ systems}\emph{\cite{2009iptps}}]
\label{thm:mmmm.Stability}
An $M/M/(M/M)$ queuing system is stable if and only if
 \beq
   \rho_c<\rho_s.
 \eeq
\end{theorem}

Note that if the Markov process $\left\{ n_{c}(t),n_{s}(t)\right\}
_{t}$ is stable, then not only it has a stationary distribution, but
also the states $(n_c=0, n_s=j), (0\leq j)$, will be visited within
finite amount of time. Practically, this means that all arriving
jobs will be served and cleared by the P2P service system in finite
time.

However in practice, job workload distribution (file length
distribution in a P2P storage system or video chunk size
distribution in a P2P streaming system), are not exponential
\cite{saroiu2002aic}. This observation motivates us to study the stability of $M/G/(M/M)$
queuing systems.

In order to do this, we consider a discrete time version of $M/G/(M/M)$ system with the number of servers bounded by $n_s\leq M_s$.
The stability condition for this system should be stronger than that of the $M/G/(M/M)$ system.
Consequently, sufficient conditions for the modified system to be stable also suffice to warranty the stability of $M/G/(M/M)$ systems.

The job-server process of such a discrete time system
is a 2-D discrete time Markov chain $(X,n_s)$ with transition probability given in Table~\ref{Tab-xnsrate}.
%

\begin{table}[!htp]
 \centering
 \caption{Transition probability of the job-server process in $M/G/(M/M)$ systems.} \label{Tab-xnsrate}
 \normalsize
\begin{tabular}{|c|c|c|}
  \hline

  From & To & Probability   \\
  \hline
  \small{$(X,n_s)$} & \small{$(X+L_i-n_s\Delta t,n_s)$} & $p_i\lambda_c\Delta t$ \\
  \small{$(X,n_s)$} & \small{$(X-n_s\Delta t,n_s+1)$}   & $\lambda_s\Delta t$ \\
  \small{$(X,n_s)$} & \small{$(X-(n_s-1)\Delta t,n_s-1)$}   & $n_s\mu_s\Delta t$  \\
  \small{$(X,n_s)$} & \small{$(X-n_s\Delta t,n_s)$}     & \footnotesize{$1-(\lambda_c+\lambda_s+n_s\mu_s)\Delta t$} \\

  \hline
  \end{tabular}
\end{table}

Note that:
\begin{enumerate}
\item It is a discrete time system with time slot interval $\Delta t$.
      We choose $\Delta t$ to be small enough so that only one event (job arrival, server arrival, server departure) happens in one slot,
      since the probability that two events happen in a single time slot is $O(\Delta t^2)$.
      When $\Delta t \rightarrow 0$, the discrete time system approaches a continuous time system.
\item We assume the distribution of job workload $X_n$ is discrete, as $$P(X_n=L_i)=p_i, \frac{L_i}{\Delta t}\in \mathbb{N}.$$
\item As a conservative approximation, when number of servers changes in a time slot,
      we use the minimal number of servers to calculate the service amount in the slot.
\item The transition probability in the table is only for the non-boundary part of the process $(X>n_s\Delta t, 0<n_s<M_s)$.
      The boundary part should be discussed independently.
\end{enumerate}

In order to find the stability condition of the system,
we use the Foster-Lyapunov criteria.
That is, we try to find a Foster-Lyapunov function and apply the following theorem.

\begin{theorem}[\textbf{Foster-Lyapunov}\emph{\cite{foster1953ams}}]\label{thm:lyapunov}
A Markov chain $\mathbf{X}(t)$ is positive recurrent, if there exists a function $V(\mathbf{X})$ that
\begin{enumerate}
\item $\forall \epsilon>0$, $\{\mathbf{X}|V(\mathbf{X})<\epsilon\}$ is a finite set.
\item $\{\mathbf{X}|\Delta V(\mathbf{X})\geq 0\}$ is a finite set,
      where $$\Delta V(\mathbf{X})=E[V(\mathbf{X}(t+1))-V(\mathbf{X}(t)) | \mathbf{X}(t)=\mathbf{X} ].$$
\end{enumerate}
Function $V(\mathbf{X})$ is called a Foster-Lyapunov function of Markov chain $\mathbf{X}(t)$ then.
\end{theorem}

The stability condition for
M/G/(M/M) systems is given in the following theorem.

\begin{theorem}[\textbf{Stability for $M/G/(M/M)$ systems}]
\label{thm:mgmm.Stability}
An $M/G/(M/M)$ queuing system is stable if
 \beq
   \rho_c<\rho_s.
 \eeq
\end{theorem}

%
%
%
%

To prove the stability of the system,
we construct a function $V(X,n_s)$ as

$$V=k\mu_cX+(n_s-\rho_s)^2+mn_s.$$

For this function $V$, it can be seen that any $k>0$ would
make $\{(X,n_s)|V(X,n_s)<\epsilon\}$ a finite set for $\forall \epsilon>0$.

We further observe that
\bda
  \Delta V &=& k(\lambda_c-n_s\mu_c)\Delta t + n_s\mu_c\mu_s\Delta t^2\\
           &+& m(\lambda_s-n_s\mu_s)\Delta t+(\lambda_s+n_s\mu_s-2\mu_s(n_s-\rho_s)^2)\Delta t.
\eda

It can be shown that as long as $\rho_c/\rho_s<1$,
there exists a tuple $(k>0,m\in \mathbb{R})$ so that $\Delta V<0$ for all $n_s \geq 0$.
For example, let
\bea
 k &=& \frac{\mu_s}{\mu_c}(2\rho_s-2\rho_c+2\frac{\rho_s}{\rho_s-\rho_c}+1), \\
 m &=& 2(\rho_s-\rho_c)-2\frac{\rho_s}{\rho_s-\rho_c}.
\eea

We have
\beq
 \frac{\Delta V}{\Delta t}=-2\mu_s(n_s-\rho_c)^2+\mu_s(\rho_c-\rho_s).
\eeq

Since $\rho_c<\rho_s$, we have $\Delta V <0$ for all $(X\geq 0,n_s\geq 0)$.

However, this transition probability is only for the non-boundary part.
Then when process is at boundary, $\frac{\Delta V}{\Delta t}$ would not equal to
\bda
  W &\triangleq& k(\lambda_c-n_s\mu_c)+ n_s\mu_c\mu_s\Delta t \\
           &+& m(\lambda_s-n_s\mu_s)+\lambda_s+n_s\mu_s-2\mu_s(n_s-\rho_s)^2
\eda

So we need to reexamine that whether $\Delta V<0$ still holds on the boundary.

\begin{enumerate}
\item $X<n_s\Delta t$ bound

On this boundary, we have $\Delta X>\lambda_c\Delta t-n_s\mu_c\Delta t$, so $\Delta V>W\Delta t$.
However, we still have that $\Delta X<\lambda_c\Delta t$, which would lead to
$$\frac{\Delta V}{\Delta t}<W+n_s\mu_c.$$

Notice here the set $\{n_s|\Delta V\geq0,n_s\geq 0\}$ is finite.
Since we have $X<n_s\Delta t$, $\{(X,n_s)|\Delta V\geq0,X\geq 0,n_s\geq 0\}$ is also finite.

\item $n_s=0$ bound

On this boundary, we still have $\Delta V=W\Delta t$.
Thus we have $\Delta V<0$.

\item $n_s=M_s$ bound

On this boundary, we have
$$\frac{\Delta V}{\Delta t}=W-\lambda_s(m+1+2(n_s-\rho_s)).$$

Thus if we choose an $M_s$ large enough, or $M_s>\rho_s-(m+1)/2$,
we would still have $\Delta V<0$ on this boundary.
If we use
$$m = 2(\rho_s-\rho_c)-2\frac{\rho_s}{\rho_s-\rho_c}$$ as mentioned,
this condition would became
$$M_s>\rho_c+\frac{\rho_s}{\rho_s-\rho_c}-\frac{1}{2}.$$
\end{enumerate}

In the summary of non-boundary and boundary part,
we have shown that function $V$ satisfies the condition that
$\{(X,n_s)|\Delta V(X,n_s)\geq 0\}$ is a finite set.
Applying Theorem \ref{thm:lyapunov}, we show that process $\left\{ X(t),n_{s}(t)\right\} _{t}$ is positive recurrent,
and thus the $M/G/(M/M)$ system is stable.


This result indicates that the relax of exponential workload assumption have no effect on the stability condition of the system.
Study on stability of more general systems (e.g. $G/G/(G/G)$) would be interesting.

\section{Study of average queue length}
\label{sec:mmmmenc}

Besides stability, another important metrics of a service system would be
the average sojourn time of jobs came into the system, which affects user experience.
A common way to study sojourn time would be looking at average queue length:
with Little's law, average queue length can be simply converted into average sojourn time.
Since study on a system with general job workload is quite difficult,
here we focus on $M/M/(M/M)$ system, where job workload is exponential.

An $M/M/(M/M)$ system can be modeled as a 2-D Markov chain
$\left\{n_c(t),n_{s}(t)\right\} _{t}$. The transition rate is listed
in Table \ref{Tab-ncnsrate}.

\begin{table}[tp]
 \centering
 \caption{Transition rate of the job-server process in $M/M/(M/M)$ systems.} \label{Tab-ncnsrate}
 \normalsize
\begin{tabular}{|c|c|c|c|}
  \hline
  From & To & Rate &  \\
  \hline
  $(n_c,n_s)$ & $(n_c+1,n_s)$ & $\lambda_c$ & for $n_c\geq 0$ \\
  $(n_c,n_s)$ & $(n_c-1,n_s)$ & $n_s\mu_c$ & for $n_c\geq 1$ \\
  $(n_c,n_s)$ & $(n_c,n_s+1)$ & $\lambda_s$ & for $n_s \geq 0$ \\
  $(n_c,n_s)$ & $(n_c,n_s-1)$ & $n_s\mu_s$ & for $n_s \geq 1$\\
  \hline
  \end{tabular}
\end{table}

Our purpose is to get the average queue length, or $E[n_c]$. The
basic method to get $E[n_c]$ is to study the balance equation of the
process, and to solve equilibrium probability
$$\pi_{ij}=P(n_s=i,n_c=j).$$ However, since this 2-D Markov process
is non time-reversible and have infinite state, it is unable to give
the close form of equilibrium distribution. However, we are able
to give the following bound.

\begin{conjecture}[\textbf{Lower bound for $M/M/(M/M)$ queue length}]
\label{thm:mmmm.lowbound}
In an $M/M/(M/M)$ queuing system, there is
 \beq
  E[n_c]>\frac{\rho_c}{\rho_s-\rho_c}.
 \eeq
\end{conjecture}

\begin{conjecture}[\textbf{Upper bound for $M/M/(M/M)$ queue length}]
\label{thm:mmmm.upbound} In an $M/M/(M/M)$ queuing system, there is
 \beq
  E[n_c]<(\frac{\mu_c}{\mu_s}+1)\frac{\rho_c}{\rho_s-\rho_c}.
 \eeq
\end{conjecture}

Note that, the lower bound $\rho_c/(\rho_s-\rho_c)$ is exactly the average queue length
of an $M/M/s$ system, where the server amount equals to $\rho_s$, which is the average server amount in $M/M/(M/M)$.
This result means that, with the same average service capacity,
a system with server dynamic would always perform worse than a static system.

When system load $\rho_c$ varies, the upper bound remains proportional with the lower bound,
with a coefficient $\mu_c/\mu_s+1$.
This result implies that, the system with higher server dynamics (which means a larger $\mu_s$) would performs
closer with a static system. This result answers the simulation result in \cite{2009iptps}.

To prove the former bounds, we need the following conjecture.

\begin{conjecture}
\label{thm:mmmm.correlation}
The correlation of $\{n_c\}$ and $\{n_s\}$ in an $M/M/(M/M)$ system is negative.
\end{conjecture}

The intuition of this conjecture is easy to understand. If the
$\{n_c\}$ and $\{n_s\}$ processes are fully decoupled, their
correlation should be zero. In $M/M/(M/M)$ system, however, the
departure rate of $n_c$ is proportional with $n_s$. So that when
$n_s$ is large, $n_c$ tend to be smaller than when $n_s$ is smaller.
That infers the negative correlation between $\{n_c\}$ and
$\{n_s\}$. However, the rigorous proof is still missing by now.
We will verify this conjecture by numerical results in the next section.

To prove the bounds of the average waiting time,
we study the generating function $F(Z,W)$ of the process,
defined as
\beq
 F(Z,W)=\sum_{i=0}^{\infty}\sum_{j=0}^{\infty}\pi_{ij}Z^iW^j.
\eeq
We further define
\bea
 H_0(W)&=&\sum_{j=0}^{\infty}\pi_{0j}W^j \\
 G_0(Z)&=&\sum_{i=0}^{\infty}\pi_{i0}Z^i
\eea
as the edge functions.
By combining definitions with the balance equation,
we know that $F(Z,W)$ satisfies the following differential equation:
\bea
\label{eqn:generating}
&&   (\mu_s+\mu_c\frac{Z}{W}-\mu_sZ-\mu_cZ)\frac{\partial F}{\partial Z}
    +\mu_c(1-\frac{1}{W})ZG_0'(Z)
\nonumber \\
&=&  (\lambda_c+\lambda_s-\lambda_cW-\lambda_sZ)F
\eea

By definition, we have a series of relationships between moments of state variables
and derivatives of generating function at the point $(Z=1,W=1)$, as follows.

\bea
  F(1,1) &=& 1\\
  \frac{\partial F}{\partial Z}(1,1) &=&   E[n_s] \\
  \frac{\partial F}{\partial W}(1,1) &=&   E[n_c] \\
  \frac{\partial^2 F}{\partial Z^2}(1,1) &=& E[n_s(n_s-1)] \\
  \frac{\partial^2 F}{\partial Z\partial W}(1,1) &=&  E[n_sn_c] \\
  G_0(1) &=&  P(n_c=0) \\
  G_0'(1) &=& E[n_s|n_c=0]P(n_c=0) \\
  G_0''(1) &=& E[n_s(n_s-1)|n_c=0]P(n_c=0)
\eea
%

Thus by applying partial derivatives to (\ref{eqn:generating}) and let $Z=1,W=1$,
we were able to get result on the moments of $n_c$ and $n_s$.

Applying $\frac{\partial}{\partial Z}$:
\beq
 E[n_s]=\rho_s
\eeq

Applying $\frac{\partial}{\partial W}$:
\beq
 G_0'(1)=E[n_s]-\rho_c
\eeq

Applying $\frac{\partial^2}{\partial Z^2}$:
\beq
 E[n_s(n_s-1)]=\rho_sE[n_s]
\eeq

Applying $\frac{\partial^2}{\partial W^2}$:
\beq \label{eqn:pwpw}
 \rho_cE[n_c]=E[n_sn_c]-E[n_s]+G_0'(1)
\eeq

Applying  $\frac{\partial^2}{\partial Z\partial W}$:
\bea \label{eqn:pzpw}
 &&\lambda_cE[n_s]+\lambda_sE[n_c]=\mu_cE[n_s(n_s-1)]+\mu_sE[n_sn_c] \nonumber \\
 &+&\mu_cE[n_s]-\mu_c(G_0'(1)+G_0''(1))
\eea

Combination based on (\ref{eqn:pwpw}) gives
\beq
  E[n_c]=\frac{\rho_c}{\rho_c-\rho_s} - \frac{\text{Cor}[n_c,n_s]}{\rho_s}.
\eeq

Since $\text{Cor}[n_c,n_s]<0$,
we have
\beq
  E[n_c]>\frac{\rho_c}{\rho_c-\rho_s}.
\eeq

Combination based on (\ref{eqn:pzpw}) gives
\beq
 E[n_c] =
 \frac{\frac{\mu_c}{\mu_s}+1}{\frac{\rho_s}{\rho_c}-1}
 +\frac{\mu_c}{\mu_s} \left(\rho_s-\frac{G_0''(1)}{\rho_s-\rho_c}\right).
\eeq

Notice that
\beq
  \frac{G_0''(1)}{\rho_s-\rho_c}
 =\frac{E[n_s(n_s-1)|n_c=0]}{E[n_s]|n_c=0}.
\eeq

Since $n_s$ and $n_c$ have a negative correlation, there is
\beq
   \frac{E[n_s(n_s-1)|n_c=0]}{E[n_s|n_c=0]}
  >\frac{E[n_s(n_s-1)]}{E[n_s]}=\rho_s,
\eeq

which gives
\beq
  E[n_c]<(\frac{\mu_c}{\mu_s}+1)\frac{\rho_c}{\rho_s-\rho_c}.
\eeq

\section{Numerical results}
\label{sec:simulation}

\begin{figure}[!htp]
  \centering
  \includegraphics[width=8cm]{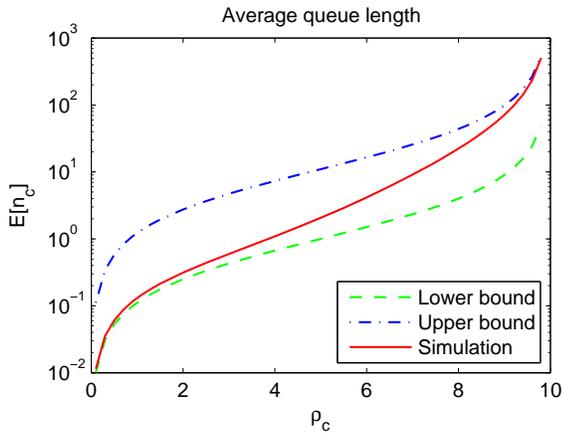}
  \caption{Average queue length: value and bound}\label{fig-simbound}
\end{figure}

We use matrix geometric analysis\cite{neuts1981mgs,latouche1999ima}
to get a numerical result of the equilibrium distribution and then
average waiting time of the system. We compare the numerical result
with the bounds we derived in the last section. The result lies in
Fig.\ref{fig-simbound}. (System parameter is $\rho_s=10$, $\mu_s=1$,
$\mu_c=10$, varying $\lambda_c$ so that $\rho_c$ changes.)

From the simulation result we see that when $\rho_c\approx0$, or the system is in light load,
the average queue length is close to the lower bound $1/(\frac{\rho_s}{\rho_c}-1)$.
While when $\rho_c\approx\rho_s$, or the system load is heavy,
the average queue length is close to the upper bound $(\frac{\mu_c}{\mu_s}+1)/(\frac{\rho_s}{\rho_c}-1)$.


%
%

\begin{figure}[!htp]

  \centering
  \includegraphics[width=8.1cm]{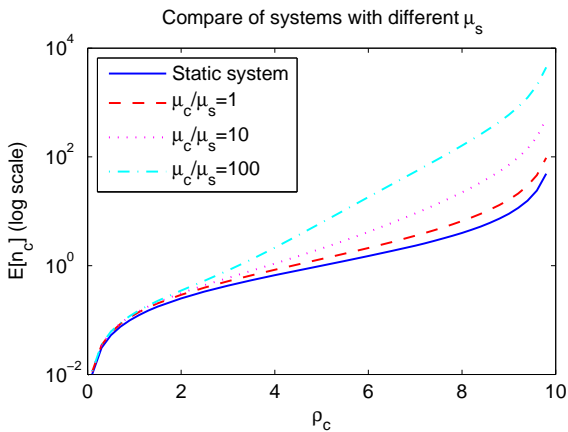}
  \caption{Compare of systems with different server dynamics}  \label{fig-dynserver}
\end{figure}

Fig.\ref{fig-dynserver}. compares systems with different server dynamics.
All the systems have the same average number of servers $\rho_s=10$, but have different $\mu_c/\mu_s$.
Result shows that with light load, all the systems perform closely.
But when the load is heavy, the average queue length is nearly proportional with $\mu_c/\mu_s+1$,
thus system with higher server dynamics will perform much better under a heavy load.

\begin{figure}[!htp]
  \centering
  \includegraphics[width=3.9cm]{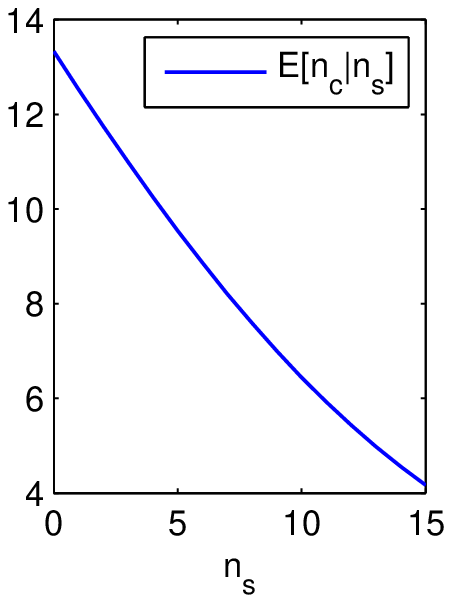}
  \includegraphics[width=3.9cm]{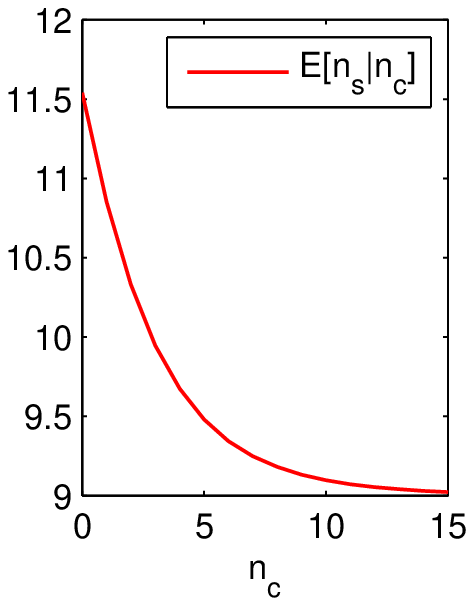}
  \caption{Correlation between server and job process}  \label{fig-cssc}
\end{figure}

We also run simulation to verify the negative correlation between $n_c$ and $n_s$ claimed in Conj.\ref{thm:mmmm.correlation}.
Conditional expectations $E[n_c|n_s]$ and $E[n_s|n_c]$ is plotted in Fig.\ref{fig-cssc}
(System parameter is $\rho_s=10, \mu_s=1,\lambda_c=8,\mu_c=1$). Note that
both curves are monotony decreasing, which indicated the negative correlation between $n_c$ and $n_s$.

\section{Conclusion}
\label{sec:concl}

In this paper we study the performance of peer-to-peer queuing
systems, where both jobs and servers arrive and depart randomly.

For $M/G/(M/M)$ models, we show that the system is stable
if its average service capacity is larger than the average workload.
This stability condition is similar to that of static service systems.

For $M/M/(M/M)$ models, we present an upper bound and a lower bound
for average queue length is given in the paper. With Little's law
these bounds can be easily converted into bounds for sojourn time.

Numerical result shows that the upper bound and lower bound are tight on heavy load and light load scenarios respectively.
For systems with same average number of servers but different server dynamics,
it is shown that they performs similarly under light load,
but system with higher server dynamics performs much better under a heavy load.

\bibliographystyle{IEEEtran}
\bibliography{IEEEabrv,p2pqueuing}

\end{document}